\documentclass[12pt]{article}
\usepackage{amsmath}
\usepackage{multicol}
\usepackage{graphicx}
\usepackage{setspace}
\usepackage{lmodern}
\usepackage{titlesec}
\usepackage{microtype}
\usepackage{multirow}
\usepackage{adjustbox}
\usepackage{fancyhdr}
\usepackage{amssymb}
\usepackage{array}
\usepackage{adjustbox}
\usepackage{float}
\usepackage{cite}

\usepackage{graphbox}
\usepackage{longtable}
\usepackage{subfigure}
\usepackage{caption}
\usepackage{booktabs}

\usepackage[section]{placeins}
\usepackage{sectsty}
\usepackage[a4paper]{geometry}
\titlespacing*{\section}{0pt}{1.1\baselineskip}{\baselineskip}
\begin{document}

		\title{\textbf{Correlation between anion related defects and ion beam induced luminescence in Y\textsubscript{4}Zr\textsubscript{3}O\textsubscript{12} }}
		\author{\normalfont Sruthi Mohan*\textsuperscript{1,2}, Gurpreet Kaur\textsuperscript{1}, Sachin Srivastava\textsuperscript{3}, \\ P. Magudapathy\textsuperscript{1}, C. David\textsuperscript{1}, G. Amarendra\textsuperscript{2}}
		\date{\textsuperscript{1} Material Science Group, Indira Gandhi Centre for Atomic Research, Kalpakkam, India.\\
		\textsuperscript{2} Homi Bhabha National Institute, Mumbai, India\\
		\textsuperscript{3} Raja Ramanna Centre for Advanced Technology, Indore, India\\
		\medskip*Corresponding author: \textit{sruthi@igcar.gov.in}}
		\maketitle
		\hrule
		\onehalfspacing
	\section*{Abstract}
\par Potential applications of Y\textsubscript{4}Zr\textsubscript{3}O\textsubscript{12} as radiation waste containment and reinforcements of Zr/Al ODS alloys vest on its irradiation stability. Fundamental studies to identify the type of defects are important in order to recognize pathways for damage alleviation. In this context, studies relating to identification of point defects and their clusters by in-situ ionoluminescence spectroscopy are taken up. The ionoluminescence spectrum acquired during 100 keV He\textsuperscript{+} ion irradiation shows two prominent bands, at 330 nm and 415 nm. Using density functional theory calculations with HSE06 hybrid exchange correlation functional, the luminescent bands have been identified to be arising due to native and irradiation induced oxygen vacancy  defects in charged and neutral configurations.

\newpage

\par The oxygen deficient fluorite compounds, pyrochlore (A\textsubscript{2}B\textsubscript{2}O\textsubscript{7}, Fd$\bar3$m), disordered fluorite (A\textsubscript{2}B\textsubscript{2}O\textsubscript{7}, Fm$\bar3$m), monoclinic pyrochlore (A\textsubscript{2}B\textsubscript{2}O\textsubscript{7}, P2\textsubscript{1}) and $\delta$-phase (A\textsubscript{4}B\textsubscript{3}O\textsubscript{12}, $R\bar3$), (A and B are rare earth transition metal species)  are capable of surviving extreme radiation environments at elevated temperatures without amorphization and hence widely recognized as potential hosts for the immobilization of high level nuclear waste \cite{weber2009materials}. Many of these oxides, especially Y\textsubscript{2}Ti\textsubscript{2}O\textsubscript{7},  Y\textsubscript{4}Zr\textsubscript{3}O\textsubscript{12} and  Y\textsubscript{4}Hf\textsubscript{3}O\textsubscript{12} are commonly found as homogeneously dispersed nano-precipitates in oxide dispersion strengthened steels containing Ti, Zr and Hf which are proposed candidate materials for structural applications in the reactor core\cite{odette2008recent,dou2014tem}. These precipitates are pivotal in strengthening the ODS steel by pinning the dislocations, trapping irradiation induced point defects and limiting the grain growth and hence the radiation response of these precipitates are directly correlated with the in-service behaviour of the steel. Both these applications of fluorite related structures in nuclear industry demand exceptionally high phase stability under severe displacement damage at elevated temperatures, which can be experimentally verified using ion beam irradiations.%, the interpretation of which often demands the understanding of native defects and evolution of them with radiation dose, dose rate, temperature, etc. 
\par The critical amorphization doses of ion-irradiated pyrochlores range from 1-100 dpa, depending on the composition. The radiation resistance is achieved by the order-to-disorder transformations to disordered fluorite structure which delays amorphization\cite{lian2002ion,wang2000ion,lian2003radiation}. The lower values of defect reaction pair energy (the sum of cation antisite formation energy and anion Frenkel pair formation energy) of $\delta$-phase compounds point to their superior radiation tolerance compared to pyrochlores\cite{mohan2020ab,sickafus2007radiation}. Existing literature on ion irradiated $\delta$- phase compounds report either a partial or full transformation from the parent rhombohedral phase to the disordered fluorite phase and delayed amorphization upon swift heavy ion irradiation and low energy ion irradiation\cite{wang2000ion,valdez2006radiation,sickafus2007radiation,tang2010order}. The critical amorphization dose up to 55 dpa has been reported in some $\delta$-phase compounds.  Nevertheless, there is no experimental or simulation study on the irradiation response of Y\textsubscript{4}Zr\textsubscript{3}O\textsubscript{12}. 

\par In general, luminescence in fluorites and related structures due to excitation from ions (Ionoluminescence - IL), electrons (Cathodoluminescence - CL) or photons (Photoluminescence - PL) are attributed to anion vacancy related defects\cite{epie2016ionoluminescence,boffelli2014oxygen,Beaumont1972aninvestigation,adair1985equilibrium}, because a strongly localized, electron-deficient vacancy readily accepts an electron and recombines radiatively.  For example, the IL spectra of CaF\textsubscript{2} mainly consist of two wide bands, explained using intrinsic anionic crystal defects and self-trapped excitons(STEs)\cite{Beaumont1972aninvestigation,williams1976time}. The cathodoluminescence of Y-doped ZrO\textsubscript{2} is reported to decrease with an increase in oxygen vacancies\cite{petrik1999laser}.  Deconvolution of CL spectra of CeO\textsubscript{2} and HfO\textsubscript{2} are arising from the radiative recombination emission caused by oxygen vacancies and there is a quenching of intensity after critical oxygen vacancy concentration\cite{thajudheen2020oxygen}. The doping of HfO\textsubscript{2} by Sc leads to a decrease of the intensity of cathodoluminescence spectra by passivating the adjacent oxygen vacancies\cite{kaichev2013xps}. The broad luminescence emissions from these structures are a result of overlapping sub-bands caused by different defect centers\cite{boffelli2014oxygen}.
\par Ion beam induced luminescence(IBIL or IL), the photon emission caused by excitation using ion beams, is an effective, yet relatively unexplored tool for gathering information about the native and irradiation-induced electronic band structure and the defect states of the target. The ionoluminescence spectra usually range from ultraviolet to infrared, with intensity proportional to the number density of defects and usually interpreted using crystal band structure and defect levels within the bandgap, similar to other luminescence techniques\cite{White1978ion,calvo2008proton,veligura2016ionoluminescence,jardin1996luminescence,perea2018ion}. In the present study, luminescence spectra produced by irradiation by light (He\textsuperscript{+}) ions in Y\textsubscript{4}Zr\textsubscript{3}O\textsubscript{12} is interpreted using hybrid functional based density functional theory calculations to understand the nature of native and irradiation induced defects in Y\textsubscript{4}Zr\textsubscript{3}O\textsubscript{12}.
\par
\begin{figure}[!hbt]
	\centering
	\includegraphics[width=0.8\linewidth]{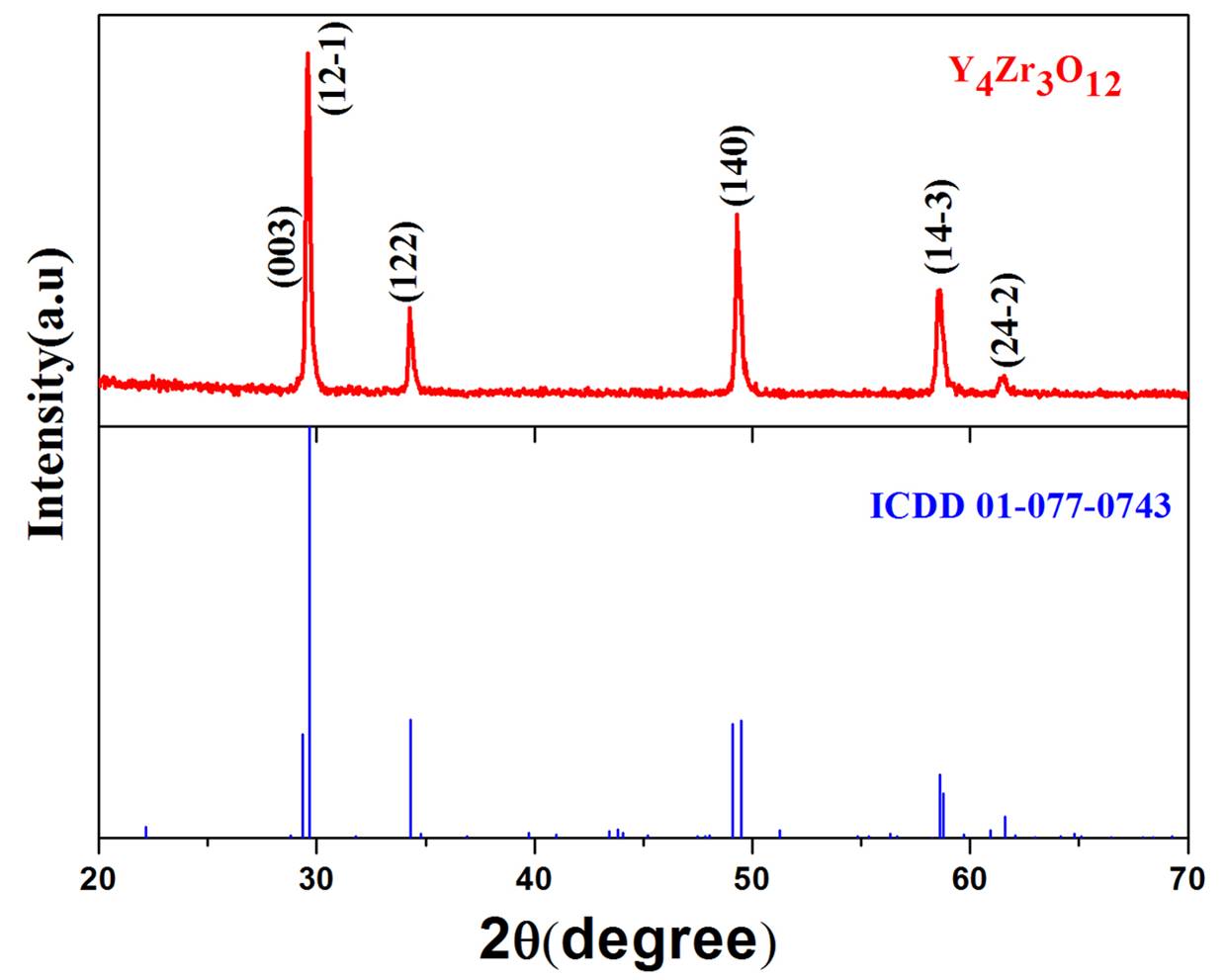}
	\caption{The XRD spectrum of as-prepared Y\textsubscript{4}Zr\textsubscript{3}O\textsubscript{12}. Corresponding ICDD data (ICDD 01-077-0743) is also shown. }
	\label{XRD}
\end{figure}

 	As the first step, the Y\textsubscript{4}Zr\textsubscript{3}O\textsubscript{12} phase is prepared from Y\textsubscript{2}O\textsubscript{3} and ZrO\textsubscript{2} precursors by ball milling (8 hours) followed by sintering (1300$^\circ$C, 120 hours). The chemical reaction involved is:
\begin{equation*}
3ZrO\textsubscript{2}+2Y\textsubscript{2}O\textsubscript{3}\rightarrow Y\textsubscript{4}Zr\textsubscript{3}O\textsubscript{12}
\end{equation*}
The phase formation was confirmed using powder-XRD given in Figure \ref{XRD}, where all the reflections can be indexed  with respect to hexagonal axes of Y\textsubscript{4}Zr\textsubscript{3}O\textsubscript{12} using ICDD data 01-077-0743 \cite{scott1977yttria,ray1977fluorite}.
 The Y\textsubscript{4}Zr\textsubscript{3}O\textsubscript{12} belongs to R$\bar{3}$ space group with two ordered structural vacancies in \textit{6c} positions along its inversion triad, [111] direction. The 3a site is occupied by Zr atom, one set of 18f positions are shared by Zr and Y atoms with a Y-occupancy of 66.6\%. Two types of oxygen, generally denoted by O\textsubscript{I} and O\textsubscript{II} are distributed in two sets of 18f positions. All the predominant reflections in Figure \ref{XRD}, (003), $(12\bar1)$, (122), (140), $(14\bar3)$ and $(24\bar2)$ are sensitive to cation ordering and the planes except (003) possess a six fold multiplicity due to the crystal symmetry.
\par
\begin{figure}[!hbt]
	\centering
	\subfigure[]{\includegraphics[width=0.80\linewidth]{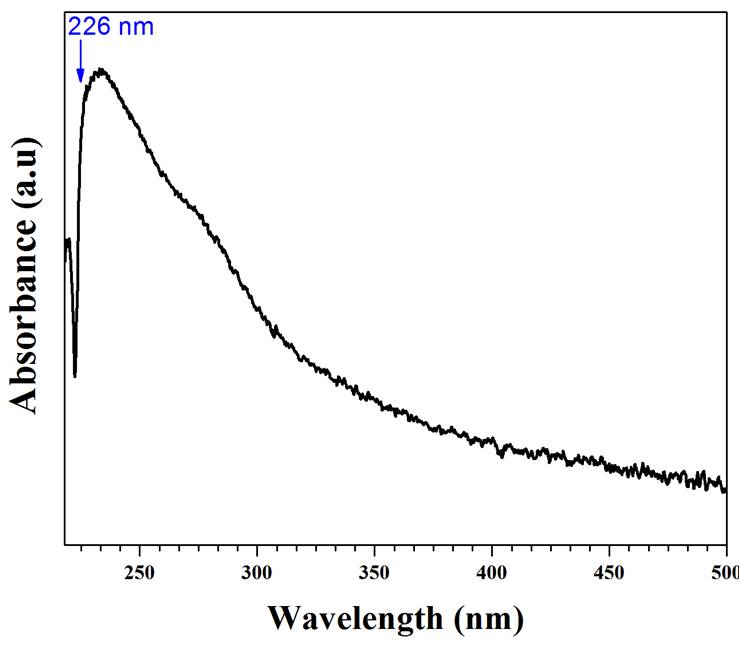}}
	\subfigure[]{\includegraphics[width=0.60\linewidth]{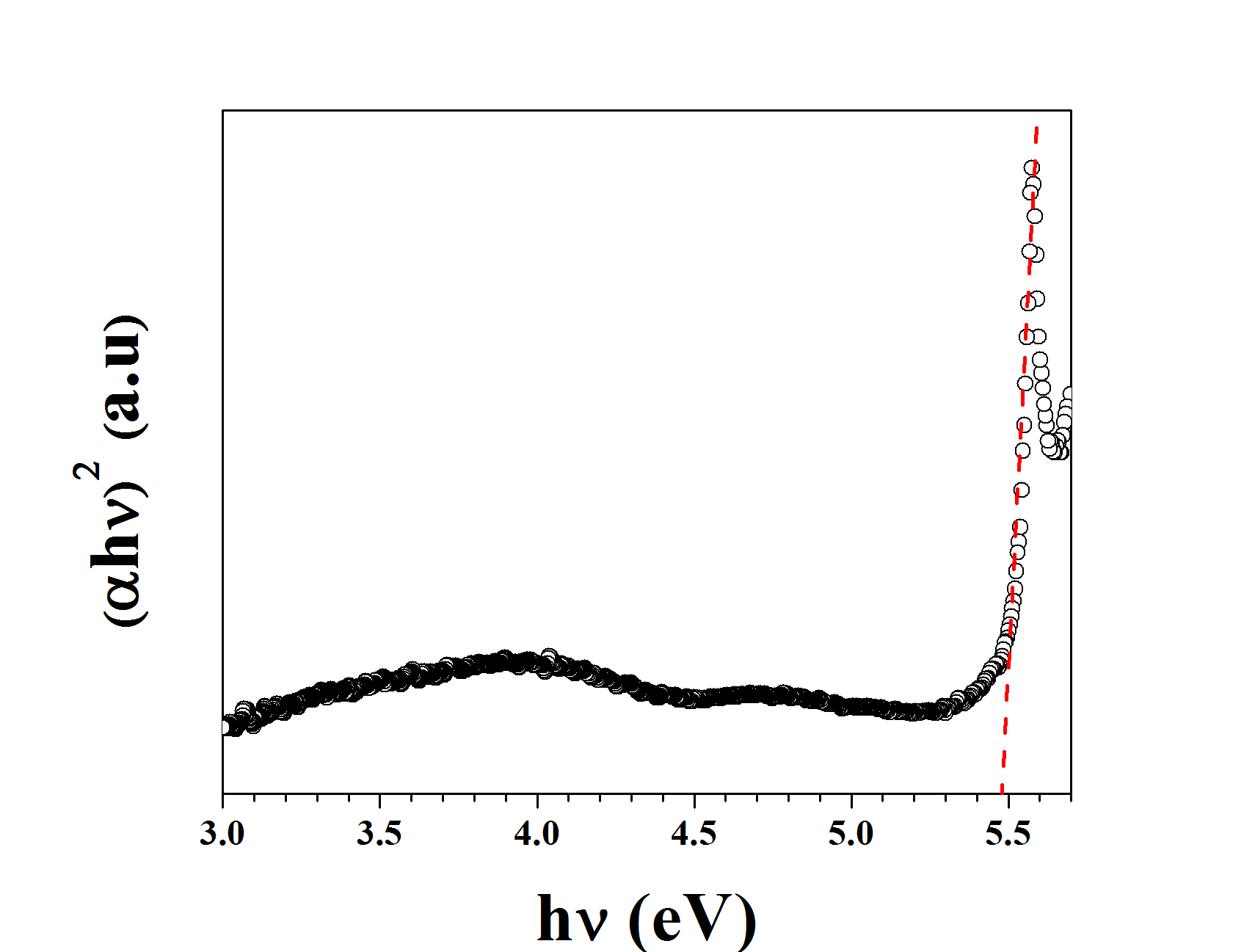}}
	\caption{(a) The UV-vis absorption spectrum of Y\textsubscript{4}Zr\textsubscript{3}O\textsubscript{12}. The absorption edge is at 226 nm. (b) The Tauc plot. The bandgap calculated from the Tauc plot is 5.47 eV.}
	\label{uv-vis}
\end{figure}

 Bandgap being a pivotal parameter in explaining any luminescence property of a material, we proceeded to measure the bandgap energy of as-prepared Y\textsubscript{4}Zr\textsubscript{3}O\textsubscript{12} using UV-vis absorption spectroscopy. The UV-visible absorption spectra and associated Tauc plot in Figure \ref{uv-vis} imply high absorption at the UV region which exponentially decreases towards the visible region. This exponential nature of decay is due to pre-existing defect states in the bandgap region\cite{rejith2019structural}. The bandgap energy value determined from the Tauc plot (Figure 2(b)) is 5.47 eV and is consistent with the earlier reported value\cite{rejith2019structural}. 
\begin{figure}[!hbt]
	\centering
	\includegraphics[width=\linewidth]{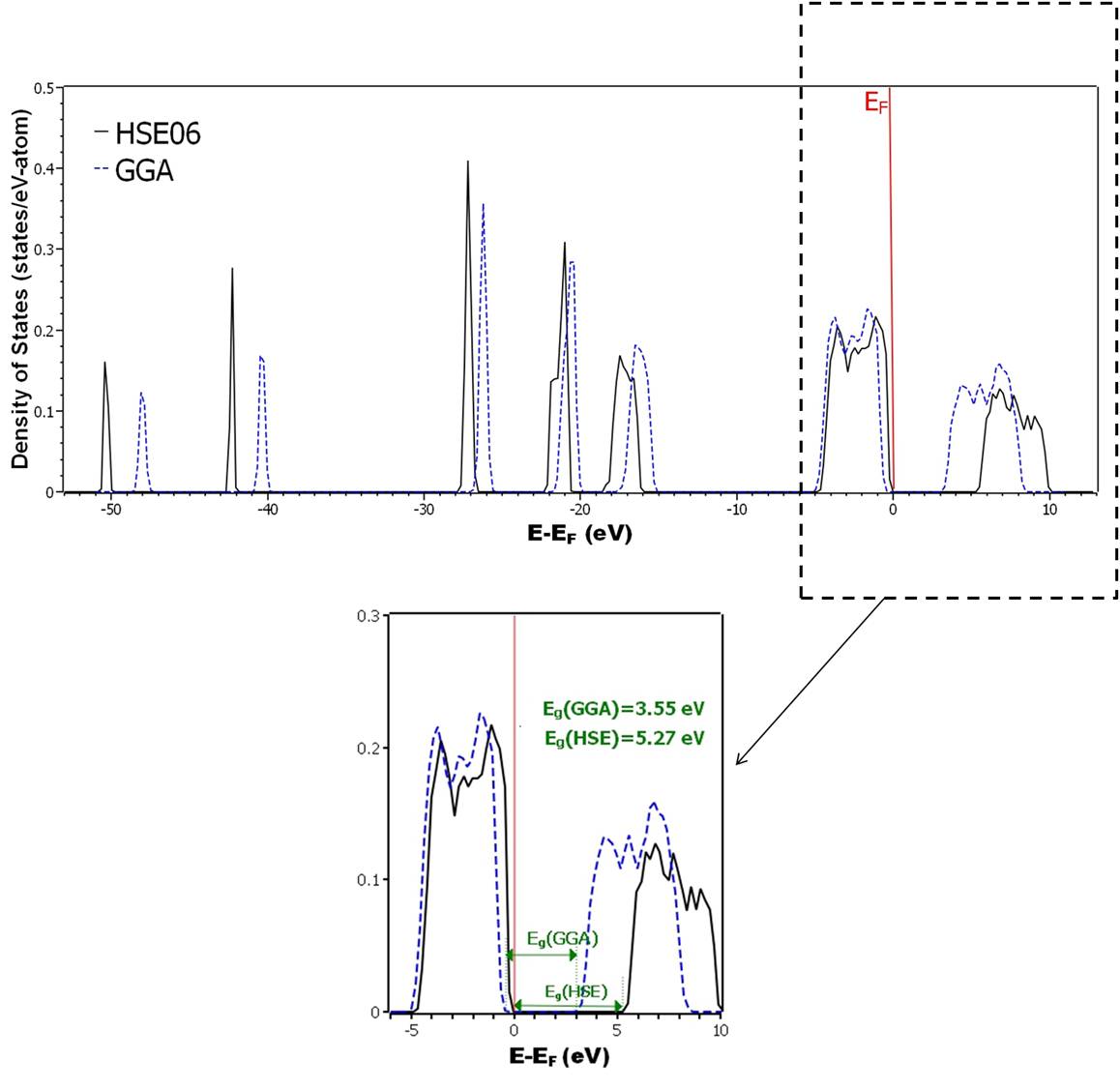}
	\caption{Total density of states (TDOS) of Y\textsubscript{4}Zr\textsubscript{3}O\textsubscript{12}, calculated using PBE-GGA(blue, dashed line) and HSE06 hybrid functional(black solid line). The red line indicates the Fermi level. The calculated bandgap is 5.27 eV for HSE06 and 3.55 eV for PBE-GGA calculations.}
	\label{HSE_DOS_Pure}
\end{figure}
\par Further DFT calculations are performed to estimate the bandgap of Y\textsubscript{4}Zr\textsubscript{3}O\textsubscript{12}. The calculations using HSE06 hybrid xc functional provide a bandgap value of 5.27 eV, which is close to the experimentally measured value while GGA calculations severely underestimate the bandgap as 3.55 eV.  Both calculations were done in a 152 atom unitcell and the HSE06 screening parameter is fixed as 0.2 {\AA}$^{-1}$. According to the total density of states plots corresponding to GGA and HSE06 calculations (shown in Figure \ref{HSE_DOS_Pure}), the conduction band minimum consists of d and f states from Y and Zr. 	
\par SRIM simulations (Figure 1 of supplementary material) have been carried out before ion irradiation experiments according to which the projected range of He$^+$ ions in Y\textsubscript{4}Zr\textsubscript{3}O\textsubscript{12} is $\sim$ 415 nm and the number of oxygen vacancies per ion-{\AA} are $\sim$ 75\% higher than the combined number of Zr and Y vacancies. Previous \textit{ab initio} molecular dynamics simulations also show that the displacement energy ($E_d$) values of cations in Y\textsubscript{4}Zr\textsubscript{3}O\textsubscript{12} is much higher than that of anions forecasting increased anionic vacancy related defect formation during irradiation\cite{mohan2020ab}.
\par The ionoluminescence spectroscopy to estimate the nature of irradiation induced defects was carried out in the \textit{in situ} IL facility at IGCAR, Kalpakkam, details of which has been recently reported\cite{srivastava2020ion}. The as-prepared Y\textsubscript{4}Zr\textsubscript{3}O\textsubscript{12} pellet is irradiated by He$^+$ ions of energy 100 keV and luminescence during irradiation is recorded at definite intervals, until maximum fluence of 1$\times$10$^{17}$ ions/cm\textsuperscript{2} is achieved at a constant current of 1 $\mu$A.

	\begin{figure}[!hbt]
	\centering
	\subfigure[]{\includegraphics[width=\linewidth]{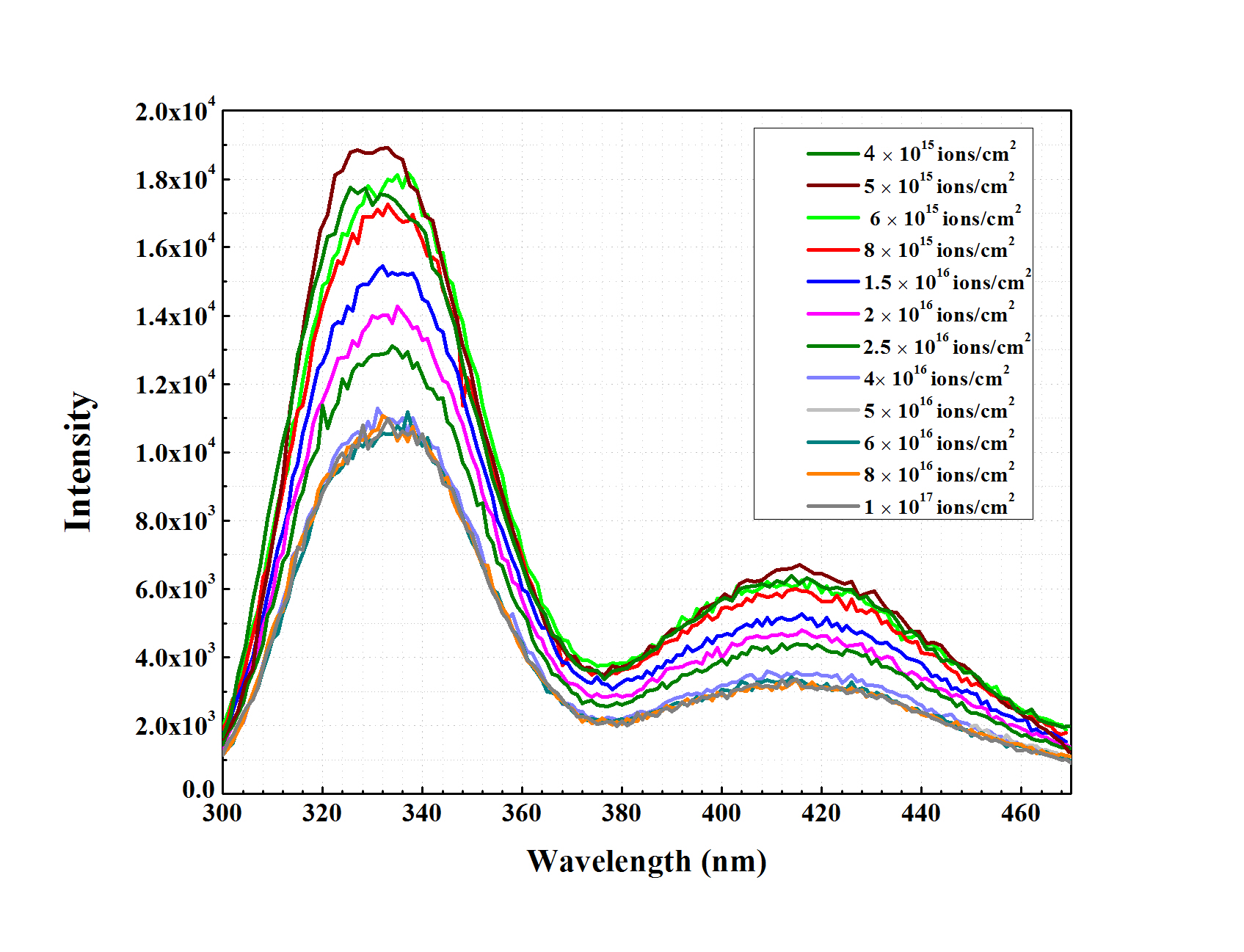}}
	\subfigure[]{\includegraphics[width=0.60\linewidth]{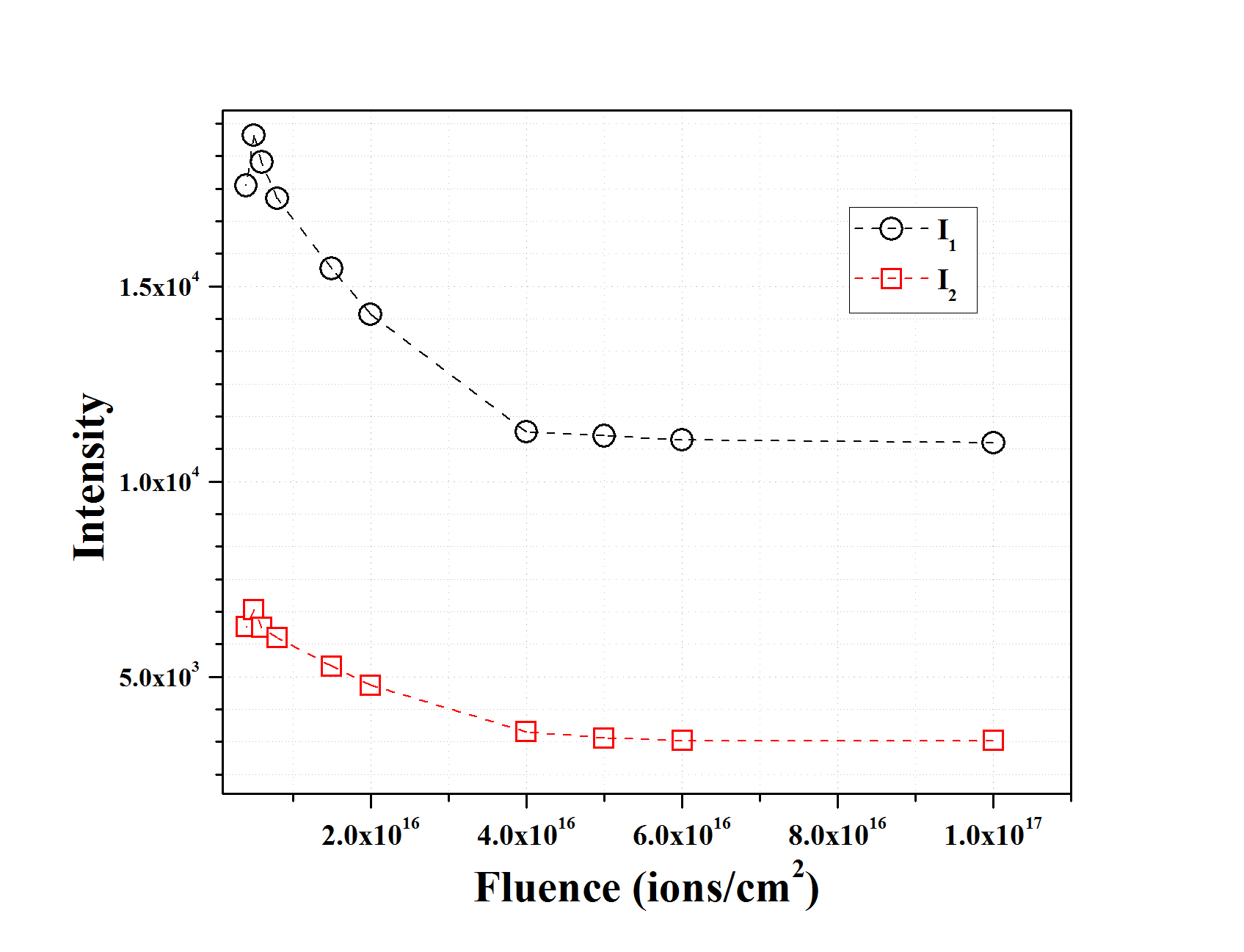}}
	%\subfigure[]{\includegraphics[width=0.45\linewidth]{IL-ratio-2.jpg}}
	\caption{(a) The ionoluminescence spectra of Y\textsubscript{4}Zr\textsubscript{3}O\textsubscript{12} acquired during irradiation with 100 keV He$^+$ ion at various fluences. The two bands observed in the spectra are marked as I\textsubscript{1} and I\textsubscript{2}.(b) Intensity variation of ionoluminescence bands with fluence. I$_1$ is the intensity of band centered at 330 nm and I$_2$ is the intensity of band centered at 415 nm. Both intensities increase initially and then decrease exponentially with fluence. %(c) The ratio of intensities of first and second peaks. %It shows that the second band is quenching faster than the first band.}
	}
	\label{IL}
\end{figure}
\par The IL spectra of  Y\textsubscript{4}Zr\textsubscript{3}O\textsubscript{12} pellet acquired for various fluences (Figure \ref{IL}) consist of two major bands centered at 330 nm (3.72 eV) and 415 nm (2.98 eV) respectively. The intensity of the first band is nearly 3 times the intensity of the second one. After an initial increase, the intensity of both bands decreases with an increase in fluence. At a fluence of $4\times10^{16}$ ions/cm$^2$, the intensities of the bands drop by ~40\% and remain the same up to $1\times10^{17}$ ions/cm$^2$. The variation of band intensities with fluence is plotted in Figure \ref{IL}(b). %The intensity of the second peak is observed to decrease at a slightly higher rate compared to the intensity of the first peak.

%	\nopagebreak 
	%\section{Role of anion-vacancies in luminescence of fluorite related structures.}
\par It is obvious that IL bands in Figure \ref{IL}(a) are far away from the band edge (5.47 eV), and therefore they can be due to deep level emission (DBE) from the native defect levels or the defect levels created by irradiation\cite{epie2016ionoluminescence}. The native and irradiation-induced defects and color-centers produce defect levels within the bandgap, the optical transitions involving which are responsible for ion beam induced luminescence\cite{veligura2016ionoluminescence,chinkov1998luminescence} and broad luminescence emissions are a result of overlapping sub-bands resulting from different defect centers\cite{boffelli2014oxygen,Beaumont1972aninvestigation,adair1985equilibrium}.	As discussed earlier, the role of oxygen vacancy related defects and quenching of luminescence intensity after critical oxygen vacancy concentration in photoluminescence, ionoluminescence and cathodoluminescence have been reported for a variety of materials\cite{jardin1996luminescence,epie2016ionoluminescence,perea2018ion,srivastava2020ion}. In relation to these literature, it is appropriate to correlate the ion beam induced luminescence spectra of Y\textsubscript{4}Zr\textsubscript{3}O\textsubscript{12} to oxygen-related defects.

\par In order to find the origin of IL peaks in Y\textsubscript{4}Zr\textsubscript{3}O\textsubscript{12}, first-principle calculations of different luminescence centers and corresponding defect levels are carried out using HSE06 hybrid functional. %The defect levels associated with \textit{3a} and \textit{18f} cation vacancies are found to be close to the valence band and hence belong to much longer wavelengths, where no IL signature is observed. 
The simplest oxygen vacancy defects created are F-centers, anion vacancies filled by 2(O$^{2-}$ or F$^{2-}$), 1(O$^-$ or F$^-$), or 0 (O-vacancy or $\alpha$-center) electrons\cite{perevalov2017oxygen}. 
\par The defect states associated with an O\textsubscript{I} vacancy creates a localized state at 3.51 eV above the valence band maximum (VBM) within the bandgap (see Figure 3 in supplementary material). In order to correlate with the experimentally observed bandgap, the defect level position can be normalized with respect to experimental bandgap as: $\frac{3.51}{5.27}\times 5.47$ eV =3.64 eV. The transition involving this defect state to valence band will give an emission at $\sim$ 340 nm.  Similarly. an O$_{II}$ vacancy in the lattice creates a level at 2.9 eV above the VBM (Figure 4 of supplementary material).  When we scale  up this to the experimentally observed bandgap value, the defect level falls at $\sim$ 3.02 eV. The transition involving this level and the valence band will be at $\sim$ 411 nm.
\par An F$^-$center is formed when either O\textsubscript{I} or O\textsubscript{II} vacancy traps an electron and it is an IL active color-center in fluorite crystals\cite{calvo2008proton}. %To examine the effect of O$^-$ vacancy in IL spectra of Y\textsubscript{4}Zr\textsubscript{3}O\textsubscript{12}, HSE06 and GGA calculations were performed. 
According to our simulations, an O\textsubscript{I}$^-$ vacancy introduces two narrow defect states in the bandgap at 3 eV (413 nm) and 4.5 (275 nm) eV respectively. When we scale up to the experimental bandgap, these defect bands will be at 3.13 eV and 4.69 eV respectively. 

\par The F$^+$ or electron deficient oxygen vacancy is responsible for the 415 nm transition in IL spectrum of sapphire\cite{jardin1996luminescence}. In Y\textsubscript{4}Z\MakeLowercase{r}\textsubscript{3}O\textsubscript{12}, an O\textsubscript{I}\textsuperscript{+} vacancy creates a %half filled
level at 3.5 eV (354 nm) above the valence band edge. Likewise, O$_{II}^+$ vacancy creates %a half filled 
defect-level at 3.46 eV (358 nm) above the valence band edge. 
The O\textsubscript{II}$^{2+}$ and O\textsubscript{II}$^{2+}$ creates defect levels at 4.2 eV (295 nm) and 4.3 eV (288 nm) respectively. 

\par
\begin{figure}[!hbt]
	\centering
	\subfigure[]{\includegraphics[width=\linewidth]{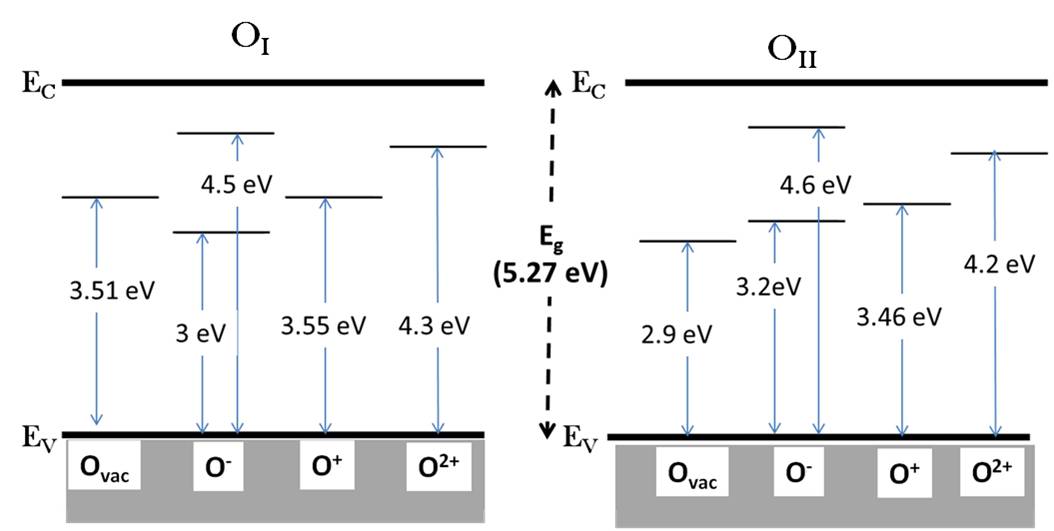}}
	%\end{figure}
	%\begin{figure}[!hbt]
	%	\centering
	\subfigure[]{\includegraphics[width=\linewidth]{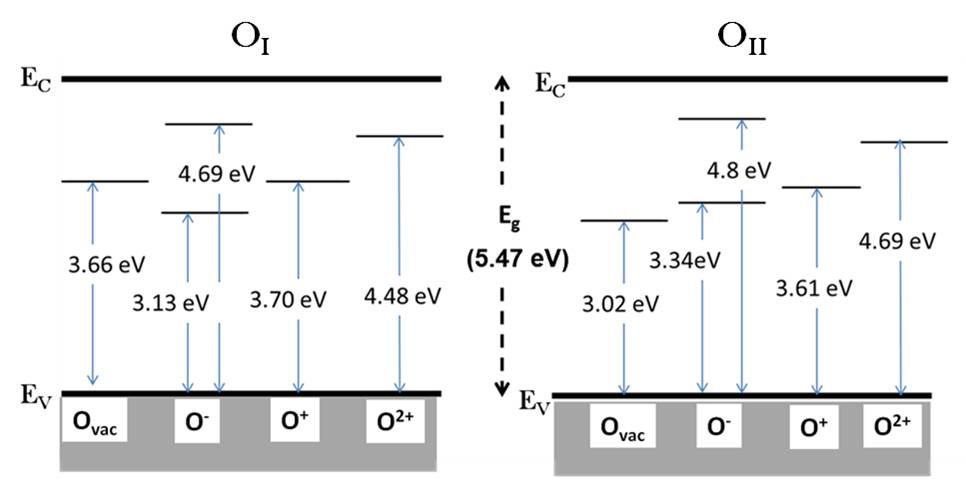}}
	\caption{(a) The defect levels in Y\textsubscript{4}Zr\textsubscript{3}O\textsubscript{12} simulated using HSE06 hybrid functional for two types of oxygen, O\textsubscript{I} and O\textsubscript{II}, in different charge states. E\textsubscript{V} and E\textsubscript{C} are valence band maximum and conduction band minimum respectively.(b) The simulated defect levels in Y\textsubscript{4}Zr\textsubscript{3}O\textsubscript{12}, scaled-up to experimentally observed bandgap, E$_g=5.47 $ eV}
	\label{band_expt}	
\end{figure}

 The defect levels in Y\textsubscript{4}Zr\textsubscript{3}O\textsubscript{12}, of O\textsubscript{I} and O\textsubscript{II} in different charge states, calculated using HSE06 functional are summarized in Figure \ref{band_expt}(a). These values are scaled up for the experimental bandgap value and shown in Figure \ref{band_expt}(b) and detailed density of states plots are provided in the supplementary document, which can be used for understanding of the contribution from different color centers towards the luminescence spectra in Figure \ref{IL}.

\par From the defect levels computed, it is evident that the energy gap between valence band and energy level positions of O\textsubscript{I} (339 nm), O\textsubscript{I}\textsuperscript{+} (335 nm) and O\textsubscript{II}\textsuperscript{+} (343 nm) vacancies fall on the wavelength range of the first band (centered at 330 nm) of IL spectra. A free electron captured by these levels will undergo recombination and simultaneous photon emission of energy $\sim$ 3.7 eV. Similarly, the radiative recombinations associated with O\textsubscript{II} (413 nm) and O\textsubscript{I}\textsuperscript{-} (371 nm) vacancies will be contributing to the second band at 415 nm. In addition to this, there might be contributions from polarons associated with the thermal activation energy of trapping of the defects. The initial increase in IL intensity with fluence is due to the irradiation induced defects. As the irradiation proceeds the native defects as well as the irradiation induced defects are getting clustered or modified\cite{wu2019evolution} into non-luminescent defects. The decrease in intensity after 6$\times$10\textsuperscript{15} ions/cm\textsuperscript{2} is due to the quenching of luminescence centers by forming defect clusters during irradiation\cite{jardin1996luminescence}. 
	 
\par In summary, the Y\textsubscript{4}Zr\textsubscript{3}O\textsubscript{12} oxide synthesized by solid-state route showed well-resolved peaks in XRD, which are in exact agreement with previously reported data. Further, the bandgap measured using UV-vis spectroscopy and hybrid-potential based electronic structure calculation are 5.27 eV and 5.47 eV respectively, which indicate it is a wide bandgap material. Here, the shortcomings of conventional DFT techniques on estimating the bandgap are compensated by selecting a suitable hybrid exchange-correlation functional. The ionoluminescence spectra acquired during 100 keV He\textsuperscript{+} ion irradiation shows two bands, centered at wavelengths 330 nm and 415 nm respectively, which can be attributed to radiative electronic transitions involving anion-related defect levels within the bandgap. The defect levels calculated using DFT-HSE06 simulations are in-line with the bands observed in the IL spectra. The O\textsubscript{I}, O\textsubscript{I}\textsuperscript{+} and O\textsubscript{II}\textsuperscript{+} vacancy defects contribute to the band at 330 nm and neutral O\textsubscript{II} and O\textsubscript{I}\textsuperscript{-} vacancies contribute to the band at 415 nm. 
\FloatBarrier

\vspace{2cm}
\section*{Acknowledgment}
\noindent The authors gratefully acknowledge Dr. Arindam Das and Dr. Binaya Kumar Sahoo, SND, IGCAR for UV-vis spectroscopic measurements.
\section*{Data Availability}
The data that support the findings of this study are available from the corresponding author upon reasonable request.
	\clearpage
\singlespacing
\bibliographystyle{aip}
\bibliography{paper}
\end{document}